\RequirePackage{ifpdf}
\ifpdf 
\documentclass[pdftex]{sigma}
\else
\documentclass{sigma}
\fi

\def\lr{{\rm L}^2({\R^d})}
\def\bra{\langle}
\def\ket{\rangle}
\def\R{\mathbb{R}}
\def\cH{\mathcal{H}}
\def\cD{\mathcal{D}}
\def\Ei{\mathrm{Ei}}
\def\dk#1#2{\frac{ d^{#2}{#1} }{ (2\pi)^{#2} }} 
\def\vp{\mathbf{p}}
\def\vq{\mathbf{q}}
\def\vy{\mathbf{y}}
\def\d{\partial}
\begin{document}

\renewcommand{\PaperNumber}{105}

\FirstPageHeading

\renewcommand{\thefootnote}{$\star$}

\ShortArticleName{Wavelet-Based Quantum Field Theory}

\ArticleName{Wavelet-Based Quantum Field Theory\footnote{This paper is a contribution to the Proceedings
of the Seventh International Conference ``Symmetry in Nonlinear
Mathematical Physics'' (June 24--30, 2007, Kyiv, Ukraine). The
full collection is available at
\href{http://www.emis.de/journals/SIGMA/symmetry2007.html}{http://www.emis.de/journals/SIGMA/symmetry2007.html}}}

\Author{Mikhail V. ALTAISKY~$^{\dag^1\dag^2}$}
\AuthorNameForHeading{M.V.~Altaisky}

\Address{$^{\dag^1}$~Joint Institute for Nuclear Research, Dubna, 141980, Russia}
\Address{$^{\dag^2}$~Space Research Institute RAS, 84/32 Profsoyuznaya Str., Moscow, 117997, Russia}
\EmailDD{\href{mailto:altaisky@mx.iki.rssi.ru}{altaisky@mx.iki.rssi.ru}}
\URLaddressDD{\url{http://lrb.jinr.ru/people/altaisky/MVAltaiskyE.html}}

\ArticleDates{Received August 15, 2007, in f\/inal form November
03, 2007; Published online November 11, 2007}

\Abstract{The Euclidean quantum f\/ield theory for the f\/ields $\phi_{\Delta x}(x)$, which depend on both the position $x$ and the resolution $\Delta x$, constructed
in \href{http://www.emis.de/journals/SIGMA/2006/Paper046/}{{\it SIGMA} {\bf 2} (2006), 046}, on the base of the
continuous wavelet transform, is considered. The Feynman diagrams
in such a~theory become f\/inite under the assumption there should be no scales in internal lines smaller than the minimal of scales of external lines.
This regularisation agrees with the existing calculations of radiative
corrections to the electron magnetic moment.
The transition from the newly constructed theory to a standard Euclidean
f\/ield theory is achieved by integration over the scale arguments.}


\Classification{42C40; 37E20}

\section{Introduction}
The description of inf\/initedimensional nonlinear systems in quantum
f\/ield theory and statistical physics always faces
the problem of divergent loop integrals emerging in the Green functions.
Dif\/ferent methods of regularisation have been applied to make the divergent integrals
f\/inite \cite{Z-J1989}. There are a few basic ideas connected with those
regularisations. First, certain minimal scale $L=\frac{2\pi}{\Lambda}$, where
$\Lambda$ is the cut-of\/f momentum, is introduced into the theory, with all
the f\/ields $\phi(x)$ being substituted by their Fourier
transforms truncated at momentum $\Lambda$:
\begin{gather}
\phi(x)\to\phi_{\left(\frac{2\pi}{\Lambda}\right)}(x) = \int_{|k|\le\Lambda} e^{-\imath k x} \tilde \phi(k) \dk{k}{d}.
\label{trunc}
\end{gather}
The physical quantities are than demanded to be independent on the
rescaling of the parameter~$\Lambda$. The second thing is the Kadanof\/f blocking
procedure~\cite{Kadanoff1966}, which averages the small-scale f\/luctuations up to a certain scale --
this makes a kind of ef\/fective interaction.

These methods are related to the self-similarity assumption:
blocks interact to each other similarly to the sub-blocks. Similarly, but
not necessarily having the same interaction strength~-- the latter can be
dependent on scale $g=g(a)$. It is the case for high energy
physics, for the developed hydrodynamic turbulence, and for many other
phenomena \cite{Vasiliev}. However there is no place for such dependence
if the f\/ields are described solely in terms of their Fourier transform~-- except for the cut-of\/f momentum. The latter representation of
the scale-dependence is rather restrictive: it determines the ef\/fective interaction of {\em all fluctuations up to a certain scale}, but says nothing
about the interaction of the f\/luctuations at a given scale \cite{AltSIGMA06}.

We have to admit that the origin of divergences is not the
singular behaviour of the interaction strength at small distance, but the
inadequate choice of the functional space used to describe these interactions.
Namely, the decomposition of the f\/ields with respect to the representations of
translation group, i.e.\  the Fourier transform
\begin{gather*}
\phi(x) = \int e^{-\imath k x} \tilde \phi(k) \dk{k}{d},
\end{gather*}
is physically sound only for the problems that clearly manifest translational invariance.
For more general
cases one can use decompositions with respect to other Lie groups, dif\/ferent from translation group
($x\to x+b$), see e.g.~\cite{IshKla1991}. The problem is what groups are physically relevant  for a f\/ield theory?
 In physical settings, along with translation invariance, the other symmetry
is observed quite often -- the symmetry with respect to scale transformations
$x\to\alpha x$. This suggests the af\/f\/ine group~\eqref{ag} may be more
adequate for self-similar phenomena than the subgroup of translations. The discrete
representation of the self-similarity idea can be found in the Kadanof\/f
spin-blocking procedure, or in application of the discrete wavelet transform
$\phi(x)= \sum d^j_k \psi^j_k(x)$ in f\/ield theory models,
considered by Battle  and Federbuch  in lattice settings \cite{Battle-book,Federbush1981}.

The decomposition with respect to the representations of af\/f\/ine group may
have a natural probabilistic interpretation.
In (Euclidean) quantum f\/ield theory the $L^2$-norm of the f\/ield $\phi(x)$ determines the
probability density of registering that particle in a certain region
$\Omega\subset \R^d$:
\begin{gather}
P(\Omega)= \int_{x\in\Omega} |\phi(x)|^2 dx, \qquad P(\R^d)=1,
\label{prd}
\end{gather}
i.e.~def\/ines a measure. The unit normalisation in \eqref{prd} is understood as
``the probability of registering a particle anywhere in space is exactly one''. This tacitly assumes
the existence of registration devices working at inf\/inite coordinate resolution.
There are no such devices in reali\-ty: even if particle is {\em there}, but its typical wavelength
is much smaller or much bigger than the typical wavelength of the measuring device there is
nonzero probability
the particle will not be registered.

For this reason it seems benef\/icial for theoretical description to use wavefunctions, or f\/ields,
that are explicitly labelled by resolution of the measuring equipment: $\phi_a(x)$. The incorporation
of an observation parameter $a$ is in excellent agreement with the Copenhagen interpretation
of quantum mechanics: $\phi_a(x)$ describes our perception of the object $\phi$ at resolution $a$,
rather than an ``object as it is'', $\phi_{a\to 0}(x)$, the existence of which is at least questionable.
Needless to say that inf\/initely small resolution ($a\to0$) requires inf\/initely high energy ($E\to\infty$)
and is therefore practically unreachable.

We suggest the normalisation for the resolution-dependent functions $\phi_a(x)$ should be
\begin{gather}
\int_{-\infty}^\infty dx \int_0^\infty d\mu(a) |\phi_a(x)|^2  =1,
\label{prs}
\end{gather}
where $\mu(a)$ is a measure of the resolution of the equipment. The normalisation \eqref{prs} will
be read as ``the probability to register the object $\phi$ anywhere in space tuning the resolution
of the equipment from zero to inf\/inity is exactly one''.

In present paper we show how the quantum f\/ield theory of scale-dependent f\/ields $\phi_a(x)$ can be
constructed using continuous wavelet transform (CWT). The integration over all scales $a$ of course will
drive us back to the standard theory. The advantage is that the Green functions
$\bra\phi_{a_1}(x_1)\cdots\phi_{a_n}(x_n)\ket$, i.e.\  those really observed in experiment,
are f\/inite -- no further renormalisation is required.

\section{Continuous wavelet transform}
Let us show how the f\/ield theory of scale-dependent f\/ields $\phi_a(x)$ can
be constructed using conti\-nuous wavelet transform \cite{Alt2002G24,AltSIGMA06}.
If $\cH$ is the Hilbert space, with is a Lie group
$G$  acting transitively on that space, and there exists a vector $\psi \in \cH$, called an
{\it admissible vector}, such that
\begin{gather*}
C_\psi = \frac{1}{\| \psi \|_{2}} \int_G |\bra \psi, U(g) \psi \ket |^2 d\mu_L(g)
<\infty, 
\end{gather*}
where $U(g)$ is a representation of $G$ in $\cH$, and $d\mu_L(g)$ is the left-invariant
measure, then for any $\phi\in \cH$ the following decomposition holds \cite{Carey1976,DM1976}:
\begin{gather}
|\phi\ket = C^{-1}_\psi \int_G |U(g) \psi \ket \bra \psi |U(g) \phi\ket d\mu_L(g),\qquad
\forall \, \phi \in \cH.
\label{dec1}
\end{gather}
The Lie group that comprises two required operations -- change of scale and translations --
is the af\/f\/ine group
\begin{gather}
x \to ax+b,\qquad \psi(x) \to U(a,b)\psi[x]=a^{-\frac{d}{2}}\psi\left(\frac{x-b}{a}\right),
\label{ag}
\end{gather}
where $x,b \in \R^d$, $a\in \R_+$.
The decomposition \eqref{dec1} with respect to af\/f\/ine group \eqref{ag} is known as
{\em continuous wavelet transform}.

To keep the scale-dependent f\/ields $\phi_a(x)$
the same physical dimension as the ordinary f\/ields $\phi(x)$ we write the coordinate
representation of wavelet transform \eqref{dec1} in $L^1$-norm \cite{Chui1992,HM1996}:
\begin{gather}
\phi(x) = \frac{1}{C_\psi} \int \frac{1}{a^d} \psi\left(\frac{x-b}{a}\right) \phi_a(b) \frac{dad^db}{a},
\label{iwt} \\
\phi_a(b) = \int \frac{1}{a^d} \overline{\psi\left(\frac{x-b}{a}\right)} \phi(x) d^dx.
\label{dwt}
\end{gather}
In the latter equations the f\/ield $\phi_a(b)$ -- the wavelet coef\/f\/icient -- has a
physical meaning of the amplitude
of the f\/ield $\phi$ measured at point $b$ using a device with an aperture $\psi$
and a tunable spatial resolution $a$.
For isotropic wavelets, which we assume in this paper, the normalisation constant
$C_\psi$ is readily evaluated using Fourier transform:
\begin{gather}
C_\psi = \int_0^\infty |\tilde\psi(ak)|^2\frac{da}{a}
= \int |\tilde\psi(k)|^2 \frac{d^dk}{S_{d}|k|}<\infty,
\label{adcf}
\end{gather}
where $S_d = \frac{2 \pi^{d/2}}{\Gamma(d/2)}$ is the area of unit sphere in $d$ dimensions.

The idea of substituting CWT (\ref{dwt}), (\ref{iwt}) into quantum mechanics or f\/ield theory is not
new \cite{HM1996,Federbush1995,HS1995,Best2000,wlb05}. However all attempts to substitute it
into f\/ield theory models were aimed to take at the f\/inal end the inverse wavelet transform
and calculate the Green functions for the ``true'' f\/ields $\bra \phi(x_1)\cdots\phi(x_n)\ket$,
i.e.~for the case of inf\/inite resolution.
Our claim is that this last step should be avoided because the inf\/inite resolution can not be
achieved experimentally. Instead we suggest to calculate the functions, which correspond to
experimentally observable f\/inite resolution correlations.
The integration over all scales $a_i$ of course will drive us
back to the standard divergent theory. The advantage of our approach is that the
Green functions $\bra\phi_{a_1}(x_1)\cdots\phi_{a_n}(x_n)\ket$ become f\/inite under certain causality assumptions.

\section{Rules of the game}
Let us start with the Euclidean f\/ield theory with the forth power
interaction $\phi^4$. The corresponding action functional can be written in the form
\begin{gather}
S_E[\phi(x)] = \frac{1}{2} \int \phi(x_1) D(x_1-x_2) \phi(x_2) dx_1 dx_2 \nonumber\\
\phantom{S_E[\phi(x)] =}{} + \frac{\lambda}{4!}
\int V(x_1,\ldots,x_4) \phi(x_1)\phi(x_2)\phi(x_3)\phi(x_4)dx_1 dx_2 dx_3 dx_4,\label{s4}
\end{gather}
where $D$ is the inverse propagator. To calculate the $n$-point Green functions of such a theory the
generation functional is constructed
\begin{gather}
\bra\phi(x_1)\cdots\phi(x_n)\ket = \left. \frac{\delta^n \ln W[J]}{\delta J^n} \right|_{J=0}, \qquad
W[J]=\int e^{-S_E[\phi] + \int J(x)\phi(x)dx}\cD\phi(x).
\label{gf}
\end{gather}
Similarly, to calculate the Green functions for scale-dependent f\/ields
$\bra\phi_{a_1}(x_1)\cdots\phi_{a_n}(x_n)\ket$ we have to construct the generating functional
for scale-dependent f\/ields $\phi_a(x)$. This is readily done by substituting wavelet transform
\eqref{iwt} into the action \eqref{s4}. This gives
\begin{gather}
W_W[J_a] =\int e^{-S_W[\phi_a] + \int J_a(x)\phi_a(x)\frac{dadx}{a}}\cD\phi_a(x),
\label{gfw}\\
\nonumber S_W[\phi_a] = \frac{1}{2}\int \phi_{a_1}(x_1) D(a_1,a_2,x_1-x_2) \phi_{a_2}(x_2)
\frac{da_1dx_1}{a_1}\frac{da_2dx_2}{a_2}  \\
\nonumber
\phantom{S_W[\phi_a] =}{} +  \frac{\lambda}{4!}
\int V_{x_1,\ldots,x_4}^{a_1,\ldots,a_4} \phi_{a_1}(x_1)\cdots\phi_{a_4}(x_4)
\frac{da_1 dx_1}{a_1} \frac{da_2 dx_2}{a_2} \frac{da_3 dx_3}{a_3} \frac{da_4 dx_4}{a_4},
\end{gather}
with $D(a_1,a_2,x_1-x_2)$ and $V_{x_1,\ldots,x_4}^{a_1,\ldots,a_4}$ denoting the wavelet images of the inverse propagator and that of the interaction potential,
respectively.

The functional \eqref{gfw} keeps the same form as its counterpart \eqref{gf} with the dif\/ference that
the functional integration over the two-argument f\/ields $\phi_a(x)$ requires their ordering in
both the position
$x$ and the scale $a$, in case the f\/ields are operator-valued. It is important that if the interaction
in the original theory \eqref{s4} is local, $V\sim \prod_{i=2}^4 \delta(x_1-x_i)$, its wavelet image
$V_{x_1,\ldots,x_4}^{a_1,\ldots,a_4}$ may be nonlocal, and vice versa.
Here the dependence of interaction on scale is only due to wavelet transform:
\[
V(x_1,\ldots,x_n) \leftrightarrow V_{x_1,\ldots,x_n}^{a_1,\ldots,a_n}.
\]
Generally speaking the explicit scale dependence of the coupling constant
$\lambda=\lambda(a)$ is also allowed. In the framework of modern f\/ield theory such dependence can not be tested: the running coupling constant
$\lambda = \lambda(2\pi/\Lambda)$, obtained by renormalisation group methods, accounts for the
collective interaction of all modes up to the certain scale $\Lambda$, but says nothing about the
interaction of modes precisely at the given scale.

The technical way to calculate the Green functions
\begin{gather*}
\bra\phi_{a_1}(x_1)\cdots\phi_{a_n}(x_n)\ket
= \left. \frac{\delta^n \ln W_W[J_a]}{\delta J_a^n} \right|_{J_a=0}
\end{gather*}
is to apply the Fourier transform to the r.h.s. of wavelet transform
\eqref{iwt} and then substitute the result
\begin{gather*}
\phi(x) = \frac{1}{C_\psi} \int_0^\infty \frac{da}{a} \int \dk{k}{d} e^{-\imath k x}
\tilde\psi(ak) \tilde \phi_a(k), 
\end{gather*}
into the action \eqref{s4}.
Doing so, we have the following modif\/ication of the Feynman diagram technique
\cite{Alt2002G24}:
\begin{itemize}\itemsep=0pt
\item each f\/ield $\tilde\phi(k)$ will be substituted by the scale component
$\tilde\phi_a(k) = \overline{\tilde\psi(ak)}\tilde\phi(k)$.
\item each integration in momentum variable will be accompanied by integration in
correspon\-ding scale variable:
\[
 \dk{k}{d} \to  \dk{k}{d} \frac{da}{a}.
 \]
\item each vertex is substituted by its wavelet transform.
\end{itemize}
For instance, for the massive scalar f\/ield propagator we have the correspondence
\[
D(k) = \frac{1}{k^2+m^2} \to  D(a_1,a_2,k) = \frac{\tilde\psi(a_1k)\tilde\psi(-a_2k)}{k^2+m^2}.
\]
Surely the integration over all scale arguments in inf\/inite limits drive us back to the
usual theory in $\R^d$, since
\[
\frac{1}{C_\psi} \int_0^\infty \frac{da}{a} |\tilde\psi(ak)|^2 = 1.
\]

In physical settings the integration should not be performed over {\em all} scales
$0\le a <\infty$.
In fact, if the system is af\/fected (prepared) at the point $x$ with the resolution $\Delta x$
and the response is measured at a point $y$ with the resolution $\Delta y$, the modes
that are essentially dif\/ferent from those two scales will hardly contribute to the result.
In the simplest case of linear propagation the result will be proportional to the product
of preparation and measuring f\/ilters
\[
\int \frac{\tilde\psi(k\Delta x)\tilde\psi(-k\Delta y)}{k^2+m^2}e^{-\imath k (x-y)} \dk{k}{d},
\]
with the maximum achieved when $\Delta x$ and $\Delta y$ are of the same order.

Because of the f\/inite resolution of measurement the causality in wavelet-based quantum f\/ield theory
\eqref{gfw} will be {\em the region causality} \cite{CC2005} in contrast to
point causality of
standard f\/ield theory. If two open balls have zero intersection $B_{\Delta x}(x) \cap B_{\Delta y}(y) = \varnothing$ the light-cone causality is applied, but if one of them is subset of another a new problem of
how to commute the part and the whole wavefunctions arises \cite{Federbush1995}. Possible solution
-- ``the coarse acts on vacuum f\/irst'' -- have been proposed in \cite{wlb05,Alt2005e}.
In fact, {\em when we perform
measurements on a quantum system of typical size $a$ we ought use system of functions with resolution
coarser or equal to $a$}: for knowing the f\/iner details requires momentum higher than $1/a$. It may seem
a trivial fact in Fourier representation: no details smaller than the radiation wavelength, used for
the experiment, can be obtained since there is insuf\/f\/icient energy for that. However in wavelet
representation this assumption should be made separately to ensure that we study any quantum system
from {\em outside} and can use only outside scales for that.

A simplest assumption of this type formulated in the language of Feynman's diagrams is:
{\em there should be no scales in internal lines smaller than the
minimal scale of all external lines}\label{caus:def}. This means that there should be no virtual particles
in internal lines unless there is suf\/f\/icient energy in external lines to excite them.

\section[Scalar field theory]{Scalar f\/ield theory}

Let us consider one-loop contribution to the two-point correlation function in
$\phi^4$-theory between two balls $B_{a_1}(x_1)$ and $B_{a_2}(x_2)$.
\begin{figure}
\centerline{\includegraphics[width=1.2in]{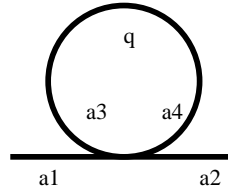}}
\caption{Tadpole diagram in scalar f\/ield scale-dependent theory with $\phi^4$-interaction.}
\label{tad4:pic}
\end{figure}
According to the above made causality statement there should be no scales in
internal loop smaller than the minimal scale of two external lines.
The value for the amputated diagram, corresponding to that shown in Fig.~\ref{tad4:pic}, is
\begin{gather}
\frac{1}{C_\psi^2} \int_{a_3,a_4 \ge A} |\tilde\psi(a_3 q)|^2 \dk{q}{d} \frac{1}{q^2+m^2}
|\tilde\psi(-a_4 q)|^2 \frac{da_3}{a_3} \frac{da_4}{a_4} ,
\label{taddef}
\end{gather}
where $A= \min (a_1,a_2)$.
In the limit of point events $A\to0$ the equation \eqref{taddef} recovers the divergent tadpole
integral ($\int \frac{1}{q^2+m^2} \dk{q}{d}$) due to normalisation \eqref{adcf}.

Let us see how the one-loop contribution \eqref{taddef} will look like for a particular
types of wavelets.
The basic wavelet $\psi$ is just an analysing function to study the object $\phi$,
and the conditions imposed on it are rather loose: practically the requirement
of normalisation \eqref{adcf} means the vanishing of the
basic wavelet Fourier image in the infra-red limit $\tilde\psi(k=0)=0$ and good localisation
properties.
For simplicity, we assert the basic wavelet $\psi$ to be isotropic and take it to be
one of the derivatives of the Gaussian, i.e.\ in Fourier space
\begin{gather}
\tilde \psi_n(k) = (-\imath k)^n e^{-k^2/2}.
\label{gnk}
\end{gather}
The normalisation constant \eqref{adcf}
can be easily evaluated for the wavelets \eqref{gnk}:
\begin{gather*}
C_{\psi_n}= \int_0^\infty (ak)^{2n} e^{-a^2k^2} \frac{da}{a} = \frac{\Gamma(n)}{2}.
\end{gather*}
Since in each internal loop there is a wavelet factor $\tilde\psi(ak)$ from
the vertex and that from the line, each internal connection to the vertex will contribute by a factor
\begin{gather*}
f(n,x) = \frac{2}{\Gamma(n)} \int_x^\infty |\tilde\psi_n(ak)|^2 \frac{da}{a}
= \frac{2}{\Gamma(n)}\int_x^\infty a^{2n-1} e^{-a^2} da,
\end{gather*}
when integrating over the scales of internal loop.
$x=Ak$ is the argument of the f\/iltering function.

\begin{figure}
\centerline{\includegraphics[width=3in]{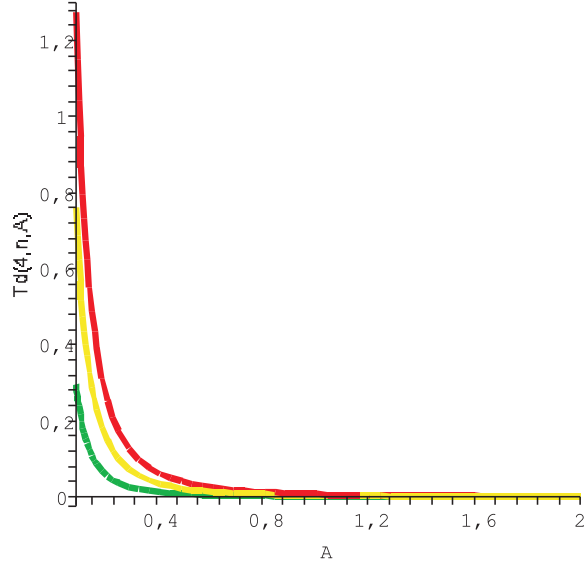}}
\caption{Scale dependence of the tadpole contributions calculated for the
f\/irst three Gaussian wavelets of the family \eqref{gnk} in $d=4$ dimensions.}
\label{tda:pic}
\end{figure}

Let us present the f\/iltering functions for the f\/irst four Gaussian wavelets \eqref{gnk} explicitly
\begin{gather*}
f(1,x) = e^{-x^2}, \\
f(2,x) = (x^2+1) e^{-x^2}, \\
f(3,x) = (x^4 +2 x^2 + 2) e^{-x^2}/2, \\
f(4,x) = (x^6 +3 x^4 + 6 x^2 + 6) e^{-x^2}/6, \\
\cdots  \cdots \cdots \cdots \cdots \cdots \cdots \cdots \cdots \cdots \cdots \cdots \cdots
\end{gather*}
Therefore, the equation \eqref{taddef}, being rewritten in dimensionless momentum units, takes the form
\begin{gather}
T^d_n(A)=\frac{S_d}{(2\pi)^d}m^{d-2} \int_0^\infty f^2(n,Amk) \frac{k^{d-1} dk}{k^2+1}.
\label{tadm1}
\end{gather}
The values of the integrals \eqref{tadm1} for the
special value of space dimension $d=4$ and wavelet numbers $n=1,2,3$ are presented below:
\begin{gather}
\label{t43}
T^4_1 = \frac{-4a^4 e^{2a^2} \Ei(1,2a^2)+2a^2}{64\pi^2a^4}m^2, \\
\nonumber T^4_2 = -\frac{\Ei(1,2a^2)e^{2a^2}a^2(4a^4 -8a^2 +4)+5a^2-2a^4-5}{64\pi^2a^2}m^2, \\
\nonumber
T^4_3 = -\frac{\Ei(1,2a^2)e^{2a^2}\!a^2(32\!+\!8a^{8}\!-\!32a^6\!+\!64a^4\!-\!64a^2) -66 +59a^2\! -42a^4\! +18a^6\!-4a^8}{512\pi^2a^2}m^2,
\end{gather}
with $a\equiv Am$ and $\Ei(1,z)=\int_1^\infty \frac{e^{-xz}}{x}dx$ being the exponential integral.
The graphs of the dependence of the values \eqref{t43} on the
dimensionless scale $a$ are shown in Fig.~\ref{tda:pic}.

\section{Theory with fermions}
The example of massive scalar f\/ield presented above demonstrates that the wavelet-based f\/ield theory of
scale-dependent functions $\phi_a(x)$ is determined by the ratio of two scales: the scale of observation
$A$ and the natural Compton scale of the theory $\frac{1}{m}$. There is a question, what will be the
result for quantum electrodynamics, the theory that comprises massive fermions and massless boson.
The answer is that localisation of photon in such a theory by any device of resolution $A$ is possible
only due to the f\/inite electron mass $m_e>0$. That is  the Compton scale is the only natural scale
in such theory.

\begin{figure}[t]
\centerline{\includegraphics[width=4cm]{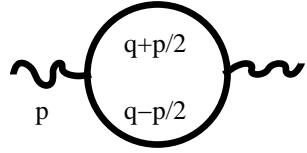}}
\caption{Polarisation operator with symmetric momenta in the loop.}
\label{vp2:pic}
\end{figure}

To illustrate this fact let us present the calculation of the vacuum polarisation diagram in
$d=4$ quantum electrodynamics (QED).
The vacuum polarisation diagram, shown in Fig.~\ref{vp2:pic} in (Euclidean) QED is given by
the following integral:
\begin{gather} \nonumber
\Pi_{\mu\nu} = -e^2 \int \dk{q}{4} \frac{{\rm Tr}\, (\gamma_\mu (\hat q+\hat p/2  -m)\gamma_\nu (\hat q -\hat p/2 - m))}{\left[(q+p/2)^2+m^2\right]\left[(q-p/2)^2+m^2\right]} \\
 \phantom{\Pi_{\mu\nu}}{} = -e^2 \int \dk{q}{4} \frac{8q_\mu q_\nu - 2 p_\mu p_\nu - \delta_{\mu\nu}(4q^2-p^2+4m^2)}{\left[(q+p/2)^2+m^2\right]\left[(q-p/2)^2+m^2\right]}
\label{pmunu_e}
\end{gather}
where the (Euclidean) identities for $\gamma$-matrices
\[
{\rm Tr}\,(\gamma_\mu \gamma_\nu) = -4 \delta_{\mu\nu}, \qquad
{\rm Tr}\,(\gamma_\mu \gamma_\alpha \gamma_\nu \gamma_\beta) = 4
(\delta_{\mu\alpha} \delta_{\nu\beta}  + \delta_{\mu\beta} \delta_{\alpha\nu}
- \delta_{\mu\nu} \delta_{\alpha\beta})
\]
were used for the evaluation of trace in the numerator of the equation~\eqref{pmunu_e}.

For def\/initeness, let us consider the f\/irst wavelet $\psi_1$ of the family \eqref{gnk}.
Each fermion line in the wavelet counterpart of the equation \eqref{pmunu_e} after integration over
internal scale variables contributes by
wavelet factor $f^2(1,x)$, where $x = A(q\pm p/2)$, for upper and lower lines in
the diagram Fig.~\ref{vp2:pic}, respectively. $A$ is the minimal scale of two
external lines.

The whole factor
\begin{gather*}
F(A) = f^2(1,A(q+p/2)) f^2(1,A(q-p/2)) = \exp\big(-A^2p^2 - 4 A^2 q^2\big)
\end{gather*}
is independent of the scalar product $pq$, and thus the resulting
equation for the vacuum polarisation in $\psi_1$ wavelet-based theory can be casted in the form
\begin{gather}
\Pi_{\mu\nu}^{(A)} = - e^2 4 \int \dk{q}{4} \exp\big(-A^2p^2 -4A^2q^2\big)
\frac{2q_\mu q_\nu - \frac{1}{2} p_\mu p_\nu +
\delta_{\mu\nu}\big(\frac{p^2}{4}-q^2-m^2\big)}{\left[(q+p/2)^2+m^2\right]\left[(q-p/2)^2+m^2\right]}.
\label{padef}
\end{gather}
Evidently the limit of inf\/inite resolution ($A\to0$) taken in equation \eqref{padef} gives the known
divergent result \eqref{pmunu_e}.

The momentum integration in equation \eqref{padef} is straightforward: having expressed all momenta in units
of electron mass $m$, we express the loop momentum in terms of the photon momentum $\vq = |\vp| \vy$
and perform the integration over the polar angle:
\begin{gather*}
\nonumber \Pi_{\mu\nu}^{(A)} = - \frac{e^2}{\pi^3} (m^2p^2) \int_0^\infty
dy y \exp\big(-A^2m^2p^2 -4A^2m^2p^2y^2\big) \int_0^\pi d\theta \sin^{2}\theta  \\
\phantom{\Pi_{\mu\nu}^{(A)} =}{} \times
\frac{2y_\mu y_\nu - \frac{1}{2} \frac{p_\mu p_\nu}{p^2} +
\delta_{\mu\nu}(\frac{1}{4}-y^2-\frac{1}{p^2})}
{\left[\frac{\frac{1}{4}+y^2+\frac{1}{p^2}}{y}+\cos\theta\right]
\left[\frac{\frac{1}{4}+y^2+\frac{1}{p^2}}{y}-\cos\theta\right],
} 
\end{gather*}
where $p$ is dimensionless, i.e.\  is expressed in units of $m$. Introducing the notation
\[
\beta(y)\equiv \frac{\frac{1}{4}+y^2+\frac{1}{p^2}}{y}
\] and using the substitution
\[
y_\mu y_\nu \to A y^2 \delta_{\mu\nu} + B y^2 \frac{p_\mu p_\nu}{p^2},
\]
under the angular integration we get
\begin{gather*}
 \Pi_{\mu\nu}^{(A)} = - \frac{e^2}{\pi^3} \big(m^2p^2\big) \int_0^\infty
dy y \exp\left(-A^2m^2p^2\big(1+4y^2\big)\right) \int_0^\pi d\theta \sin^{2}\theta \times \\
\phantom{\Pi_{\mu\nu}^{(A)} =}{}\times
\frac{
\delta_{\mu\nu}\big( (2A-1)y^2 + \frac{1}{4}-\frac{1}{p^2}\big) + \frac{p_\mu p_\nu}{p^2} \big(2By^2-\frac{1}{2}\big)
}
{\beta^2(y)-\cos^2\theta},
\end{gather*}
where $A$ and $B$ depend only on the modulus of $y$, but not on the direction, and can be expressed
in terms of angle integrals
\begin{gather*}
\nonumber I_k(y) \equiv \int_0^\pi d\theta \frac{\sin^{2}\theta \cos^{2k}\theta}{\beta^2(y)-\cos^2\theta}, \\
          I_0(y) = \pi (1-\sqrt{1 - \beta^{-2}(y)}),\\ 
\nonumber I_1(y) = -\frac{\pi}{2} + \beta^2(y)I_0(y), \\
\nonumber        \cdots \cdots \cdots \cdots\cdots\cdots\cdots\cdots
\end{gather*}
so that
$ 4A+B=1,\quad A+B=I_1/I_0$, from where we get
\[
A = \frac{1}{3} + \frac{\pi}{6} I_0^{-1}(y) -\frac{1}{3}\beta^2(y), \qquad
B = -\frac{1}{3} -\frac{2\pi}{3} I_0^{-1}(y)+\frac{4}{3}\beta^2(y).
\]
Finally, writing the polarisation operator as a sum of transversal and longitudinal
parts, we have the equations
\begin{gather} \nonumber
\Pi_{\mu\nu}^{(A)} \equiv \delta_{\mu\nu} \pi^{(A)}_T + \frac{p_\mu p_\nu}{p^2} \pi^{(A)}_L,\\
\pi^{(A)}_T = -\frac{e^2}{3\pi^2} m^2p^2 \int_0^\infty dy y \exp\left(-A^2m^2p^2\big(1+4y^2\big)\right)
\label{piat} \\ \nonumber
{}\times \!\left[y^2\! +  \!\left(\!1-\!\sqrt{\frac{\frac{1}{16}+y^4+\frac{1}{p^4}-\frac{y^2}{2}+\frac{1}{2p^2}+\frac{2y^2}{p^2}}
{\!\!\!\left(\frac{1}{4} + y^2 + \frac{1}{p^2}\right)^2}} \right)
\left(\frac{5}{8}-\frac{4}{p^2}-\frac{2}{p^4}-2y^2\left(1+\frac{2}{p^2}\right)\!-2y^4\right)\!\right]\!,
\\
\pi^{(A)}_L = -\frac{e^2}{3\pi^2} m^2p^2 \int_0^\infty dy y \exp\left(-A^2m^2p^2\big(1+4y^2\big)\right)
\label{pial} \\ \nonumber
{}\times\!\left[-4y^2 \!+  \!\left(\!1-\!\sqrt{\frac{\frac{1}{16}+y^4\!+\frac{1}{p^4}-\frac{y^2}{2}\!+\frac{1}{2p^2} +\frac{2y^2}{p^2}}
{\!\!\!\left(\frac{1}{4} + y^2 + \frac{1}{p^2}\right)^2}} \right)
\left(8y^4+2y^2\!\left(1+\frac{8}{p^2}\right)\!+\frac{4}{p^2}+\frac{8}{p^4}-1\right)\!\right]\!.\!
\end{gather}
\begin{figure}
\centering \includegraphics[width=3in]{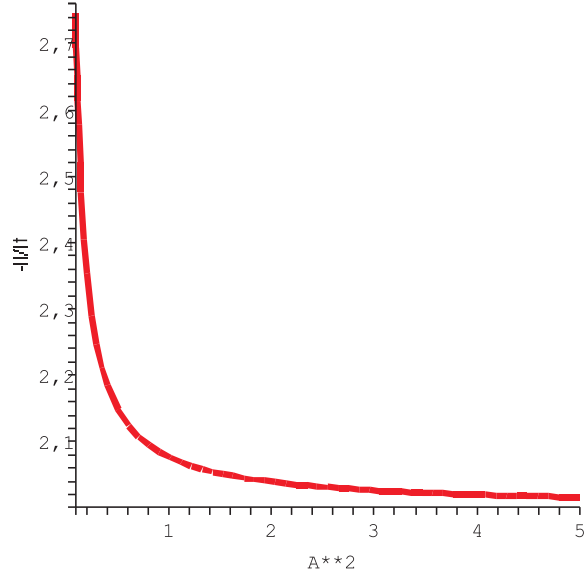}
\caption{The ratio of the longitudinal
part to the transversal part of the polarisation operator $-\pi^{(a)}_L/\pi^{(a)}_T$, shown for $p=5.0$.}
\label{IlIt:pic}
\end{figure}

The integrals (\ref{piat}), (\ref{pial}) can be evaluated in the limiting case $p^2\gg1$, when the external momentum
is much greater the electron mass. In this case
\begin{gather*}
\pi^{(A)}_T = -\frac{e^2}{6\pi^2} m^2p^2 \int_0^\infty \!\! dt \exp\left(-A^2m^2p^2(1+4t)\right)
\!\!\left[t + \left(1-\!\sqrt\frac{(\frac{1}{4}-t)^2}{(\frac{1}{4}+t)^2 }\right)
\left(\frac{5}{8}-2t -2t^2\right) \right]\!, \\
\pi^{(A)}_L = -\frac{e^2}{6\pi^2} m^2p^2 \int_0^\infty \!\! dt \exp\left(-A^2m^2p^2(1+4t)\right)
\!\!\left[-4t + \left(1-\!\sqrt\frac{(\frac{1}{4}-t)^2}{(\frac{1}{4}+t)^2 }\right)
\big(8t^2+2t-1\big) \right]\!,\!
\end{gather*} and can be evaluated as a sum $\int_0^\infty = \int_0^{1/4} + \int_{1/4}^\infty$.
This gives
\begin{gather*}
\nonumber \pi^{(a)}_T = -\frac{e^2}{6\pi^2} p^2
\left\{
\frac{e^{-a^2p^2}}{8a^6p^6}\big(4a^4p^4-a^2p^2-1\big)
+\frac{e^{-2a^2p^2}}{8a^6p^6}\big(-4a^4p^4+2a^2p^2+1\big)\right. \\
\left. \phantom{\pi^{(a)}_T =}{} -\frac{1}{2}\Ei\big(1,a^2p^2\big) + \Ei\big(1,2a^2p^2\big) 
\right\}, \\
\nonumber \pi^{(a)}_L = -\frac{e^2}{6\pi^2} p^2
\left\{
\frac{e^{-a^2p^2}}{4a^6p^6}\big(-2a^4p^4-a^2p^2+2\big)
+\frac{e^{-2a^2p^2}}{4a^6p^6}\big(2a^4p^4-a^2p^2-2\big)\right. \\
\left. \phantom{\pi^{(a)}_L =}{} +\frac{1}{2}\Ei\big(1,a^2p^2\big) - \Ei\big(1,2a^2p^2\big)
\right\}, \qquad a=Am.
\end{gather*}
In the limiting case of $a^2\to\infty$ the ratio of the longitudinal
part to the transversal part $-\pi^{(a)}_L/\pi^{(a)}_T\to2$, see Fig.~\ref{IlIt:pic}.

\section{Relation to the usual regularisations}
The decomposition of wavefunctions with respect to representation of the
af\/f\/ine group is of course a basis for certain regularisation, but is not
identical to known regularisations, such as the Wilson RG procedure \cite{WK1974,Wilson1983}, see \cite{AltSIGMA06} for more details. In the Wilson renormalisation group
the integration over a thin shell in momentum space
$[\Lambda e^{-\delta l},\Lambda)$ averages the fast modes into the ef\/fective
slow modes. The ef\/fective coupling constant $\tilde g(\Lambda)$ in such a theory
stands for the ef\/fective interaction of modes with $k\le\Lambda$, rather than
being a coupling constant describing the interaction strength at a given scale.

The renormalisation group (RG), that makes use of substitution of initial f\/ields $\phi(x)\in\lr$
by the scale-truncated f\/ields \eqref{trunc} makes the coupling constants dependent on the cut-of\/f momentum $\Lambda$, and requires
that the f\/inal physical results should be independent
of the introduced scale
\[
\Lambda\d_{\Lambda}(\text{Physical quantities})=0.
\]
The standard regularisation schemes, the Wilson RG, the Pauli--Villars
regularisation, etc., share an important common feature: if the
studied process has a typical observation scale~-- the inverse momentum
of external lines,~-- then the smaller scale contributions are ef\/fectively
suppressed by a regularisation parameter (cut-of\/f momentum, large mass,~etc.),
with their averaged ef\/fect being incorporated into the observable
scale parameters.

Let us illustrate this using the example of vertex diagram in QED,
and show that the wavelet transform with the above proposed causality
assumption acts similarly.

\begin{figure}
\centerline{\includegraphics[width=4in]{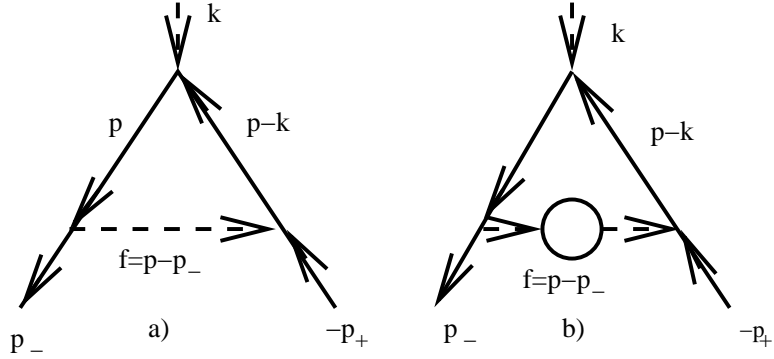}}
\caption{First and second radiation correction to the electron
magnetic moment. For the second radiation corrections only one of
the diagrams is shown.}
\label{rc:pic}
\end{figure}

The equation for the anomalous magnetic moment of the electron
\begin{gather}
\mu = \frac{e\hbar}{2mc}\left(1+\frac{\alpha}{2\pi}
-0.328\frac{\alpha^2}{\pi^2}\right),
\label{emm}
\end{gather}
where $\alpha=\frac{e^2}{\hbar c}$ is the f\/ine structure constant,
provides the basis for the most precise tests of quantum electrodynamics.
The unit term in the equation \eqref{emm} is just a magnetic moment of
the electron, the second is the f\/irst radiation correction, corresponding
to the diagram shown in Fig.~\ref{rc:pic}a, f\/irst calculated by Schwinger, the
second term, corresponding
to the diagram shown in Fig.~\ref{rc:pic}b, was f\/irst
calculated by C.~Sommerf\/ield.
The calculation is based on the evaluation of the electron
formfactor
\begin{gather*}
j^\mu = \bar u_2 \Gamma^\mu u_1, \qquad
\Gamma^\mu = \gamma^\mu f(k^2)-\frac{1}{2m}g(k^2) \sigma^{\mu\nu}k_\nu
\end{gather*}
Following \cite{LL4} we present the limitations on internal line momenta in the f\/irst radiation
correction,  Fig.~\ref{rc:pic}a,~-- for the second one, shown in Fig.~\ref{rc:pic}b,
the procedure is the same.

The matrix element corresponding to the electron current shown in
Fig.~\ref{rc:pic}a is given by
\begin{gather*}
-\imath e \bar u(p_-) \Gamma^\mu u(-p_+) = (-\imath e)^3 \bar u(p_-)
\gamma^\nu \imath \int G(p)\gamma^\mu G(p-k)\gamma^\lambda D_{\lambda\nu}(f)
u(-p_+) \dk{p}{4},
\end{gather*}
or explicitly
\[
\bar u(p_-)\left(
\gamma^\mu f(k^2)-\frac{1}{2m}g(k^2) \sigma^{\mu\nu}k_\nu
\right) u(-p_+) = \imath \int \frac{\bar u(p_-)\phi^\mu(p) u(-p_+) d^4p}{
(p^2-m^2)[(p-k)^2-m^2]},
\]
where
\[
\phi^\mu(p) = -e^2 \frac{\gamma^\nu (\hat p+m)\gamma^\mu (\hat p - \hat k+m)\gamma_\nu}{4\pi^3 (p_--p)^2}.
\]
The loop integration is performed in momentum $f=p-p_-$ instead of $p$,
so that
\begin{gather*}
f^2=(p-p_-)^2=-2{\bf p}^2(1-\cos\theta) = - \frac{t-4m^2}{2}(1-\cos\theta), \qquad
\theta = \angle ({\bf p},{\bf p_-})
\end{gather*}
 and leads to the integrals
\begin{gather*}
(I,I^\mu,I^{\mu\nu}) = \int \frac{(1,f^\mu,f^\mu f^\nu)}{1-\cos\theta} \frac{d\Omega_f}{2\pi}.
\end{gather*}
These integrals have infra-red divergences of the form
\begin{gather}
I = \int_0^{t-4m^2} \frac{d{\bf f}^2}{{\bf f}^2},
\label{X}
\end{gather}
where $t=k^2$. The regularisation is performed by introducing the small but f\/inite photon
mass ($\lambda\ll m$) and corresponding shift of the momentum $f^2\to f^2-\lambda^2$.
Analogous consideration can be presented for the integrals $I^\mu$ and $I^{\mu\nu}$.

Thus, in the f\/inal limit of the large scale magnetic f\/ield ($k \to 0$)
the integration in \eqref{X} is performed over the momenta less or equal
than $k^2-4m^2$, i.e.\  in the scales larger than the scale of external
lines. Similar consideration can be applied to other diagrams of radiation
corrections, including that shown in Fig.~\ref{rc:pic}b. This exactly corresponds to the idea presented above in this paper on page~\pageref{caus:def} in terms of continuous wavelet transform.

\section{Conclusion}
In this paper we sketched a way of constructing quantum f\/ield theory for the f\/ields that depend
on both the position and the scale using the continuous wavelet transform.
The continuous wavelet transform  has been already used for regularisation
of f\/ield theory models \cite{Federbush1995,Battle-book,Best2000}. The novelty of present approach
(see also \cite{AltSIGMA06}), consists in understanding the scale-dependent f\/ields $\phi_a(x)$ --
the wavelet coef\/f\/icients~-- as physical amplitudes of the f\/ields, measured at a given resolution $a$.
This seems to be advantageous if compared to mere regularisation, which is to be considered at the
limit $a\to0$ in the f\/inal results. The advantage is in explicit equations for the correlation between
f\/ields of dif\/ferent scales $a_i$, allowed at the same location $x$. Such correlations do really take
place in the process of quantum measurement, when the system is initially measured at large scale, and
then on a small scale, -- say the measurement of the angular momentum of a molecule followed by a
measurement of an electron angular momentum. Technically, the restriction of minimal scale of
all internal lines in a (wavelet) Feynman diagram by the minimal scale of external lines provides
the absence of processes with energies not supplied by the experimental device or the environment.
This limitation makes the theory free of ultra-violet divergences.

Doing so we obtain a nonlocal f\/ield theory with region causality \cite{CC2005,Alt2005e} instead
of point causality, accompanied by corresponding problems of nonlocal f\/ield theory \cite{AE1974,Efimov1985}.
This makes the wavelet approach attractive for further applications in high energy physics and
condensed matter f\/ield theoretic models. To go further in this direction we need to elucidate
the ef\/fects of gauge invariance to the multiscale decomposition, but this will be the subject of the
subsequent paper.

\subsection*{Acknowledgements}

The author is thankful to Professor N.V.~Antonov for critical reading of the
manuscript.

\pdfbookmark[1]{References}{ref}
\LastPageEnding

\end{document}